\documentclass[submission, Phys]{SciPost}

\usepackage{amsmath}
\usepackage{amsbsy}
\usepackage{bm}

\hypersetup{
    colorlinks,
    linkcolor={red!50!black},
    citecolor={blue!50!black},
    urlcolor={blue!80!black}
}

\usepackage[bitstream-charter]{mathdesign}
\DeclareSymbolFont{usualmathcal}{OMS}{cmsy}{m}{n}
\DeclareSymbolFontAlphabet{\mathcal}{usualmathcal}

\newcommand{\iu}{{i\mkern1mu}}

\begin{document}
\begin{center}{\Large \textbf{
Skyrmion crystals in the triangular Kondo lattice model\\
}}\end{center}

% TODO: write the author list here. Use first name (+ other initials) + surname format.
% Separate subsequent authors by a comma, omit comma and use "and" for the last author.
% Mark the corresponding author with a superscript star.
\begin{center}
Zhentao Wang\textsuperscript{1,2,*} and
Cristian D. Batista\textsuperscript{1,3}
\end{center}

\begin{center}
{\bf 1} Department of Physics and Astronomy, The University of Tennessee, Knoxville, Tennessee 37996, USA
\\
{\bf 2} School of Physics and Astronomy, University of Minnesota, Minneapolis, Minnesota 55455, USA
\\
{\bf 3} Quantum Condensed Matter Division and Shull-Wollan Center, Oak Ridge National Laboratory, Oak Ridge, Tennessee 37831, USA
\\
{\bf *} Present address: Center for Correlated Matter and School of Physics, Zhejiang University, Hangzhou 310058, China
\end{center}

\begin{center}
\today
\end{center}

\section*{Abstract}
{\bf
We present a systematic study of the formation of skyrmion crystals in a triangular Kondo Lattice model for generic electron filling fractions. Our results indicate that the four-sublattice chiral antiferromagnetic ordering that was reported more than one decade ago can be understood as the dense limit of a sequence of skyrmion crystals whose lattice parameter is dictated by the Fermi wave-vector. 
%Moreover, the skyrmion crystals remain stable at zero temperature even in absence of easy-axis anisotropy when the Kondo exchange becomes   larger than a critical value.  
This observation has important implications for the ongoing search of skyrmion crystals in metallic materials with localized magnetic moments.
}

\vspace{10pt}
\noindent\rule{\textwidth}{1pt}
\tableofcontents\thispagestyle{fancy}
\noindent\rule{\textwidth}{1pt}
\vspace{10pt}

\section{Introduction}
%\label{Intro}

The  magnetic skyrmion crystals (SkX) envisioned by Bogdanov and Yablonskii~\cite{BogdanovAN1989,RosslerUK2006} were finally discovered in 2009  studying a family of chiral magnets that includes MnSi, Fe$_{1-x}$Co$_x$Si, FeGe and Cu$_2$OSeO$_3$~\cite{MuhlbauerS2009,YuXZ2010,YuXZ2011,SekiS2012,AdamsT2012}. These SkX  arise as a superposition of three magnetic spirals produced by competition between ferromagnetic exchange and Dzyaloshinskii-Moriya (DM) interaction~\cite{Dzyaloshinsky1958,MoriyaT1960,MoriyaT1960a} induced by the non-centrosymmetric structure of these materials. More recently, SkX were also reported in centrosymmetric materials, such as BaFe$_{12-x-0.05}$Sc$_x$Mg$_{0.05}$O$_{19}$, La$_{2-2x}$Sr$_{1+2x}$Mn$_2$O$_7$, Gd$_2$PdSi$_3$, Gd$_3$Ru$_4$Al$_{12}$, GdRu$_2$Si$_2$, NiI$_2$, Mn$_4$Ga$_2$Sn, and EuAl$_4$~\cite{YuX2012,YuXZ2014,MallikR1998a,SahaSR1999,KurumajiT2019,ChandragiriV2016,HirschbergerM2019,KhanhND2020,AmorosoD2020,ChakrabarttyD2022,ShangT2021,KanekoK2021,TakagiR2022}. In these cases, the underlying spiral structure arises from competition between different exchange couplings or dipolar interactions~\cite{OkuboT2012,LeonovAO2015,LinSZ2016,Hayami16,BatistaCD2016_review}. 
%required to produce SkX 

To date, most magnetic SkX have been reported in metals, where the magnetic textures act as an effective potential for conduction electrons and reconstruct the electronic bands. Moreover, in the adiabatic limit and in absence of spin-orbit coupling~\cite{Zhang20}, the skyrmion density  turns out to be proportional to a fictitious magnetic field that couples to the orbital degrees of freedom of the conduction electrons enabling 
novel response functions, such as the well-known topological Hall effect~\cite{OnodaM2004,YiSD2009,HamamotoK2015,GobelB2017} and the current-induced skyrmion motion~\cite{JonietzF2010,YuXZ2012,SchulzT2012,NagaosaN2013}.  Hall conductivities comparable to the quantized value ($e^2/h$)  can in principle be achieved if the ordering wave vector of the SkX is comparable to the Fermi wave vector $k_F$. As it was shown in a recent work~\cite{WangZ2020},  this condition can be naturally fulfilled in $f$-electron materials where localized magnetic moments interact via Kondo exchange with conduction electrons. In the weak-coupling limit of the Kondo Lattice model (KLM) that is traditionally used to describe these materials, the localized moments interact via an effective Ruderman-Kittel-Kasuya-Yosida (RKKY) interaction mediated by the conduction electrons~\cite{RudermanMA1954,KasuyaT1956,YosidaK1957}. Since the wave length of the  SkX induced by this interaction (in combination with a small easy-axis anisotropy) is $\pi/k_F$,  the resulting SkX can produce a very large Hall conductivity (of order $e^2/h$)~\cite{WangZ2020}.

%Our results are potentially relevant for the rare earth based materials Gd$_2$PdSi$_3$ and Gd$_3$Ru$_4$Al$_{12}$ that contain a magnetic field induced SkX phase in their phase diagrams~\cite{Mallik1998_paramana,Saha1999,Kurumaji2019,ChandragiriV2016,Hirschberger2018}.

One year before the remarkable experimental discovery of SkX, a  four-sublattice chiral antiferromagnetic (AFM) ordering was reported to be present in the zero-temperature phase diagram of the triangular KLM (TKLM)~\cite{MartinI2008}. This chiral ordering with uniform spin chirality can be regarded as the dense (short wavelength) limit of SkX with two skyrmions per magnetic unit cell~\cite{park2023}. 
%sequence of SkX that emerge in the TKLM for general filling fractions. 
%In the dense limit, 
%each magnetic unit cell
%of the four-sublattice chiral AFM ordering
%contains two
%magnetic
%skyrmions and 
%the scalar spin chirality
%or skyrmion density
%becomes  uniform.
%(the spins of each triangular plaquette span a quarter of the solid angle of the sphere).
As it was pointed out in previous works~\cite{MartinI2008,AkagiY2012}, the  triple-$\bm{Q}$
%($\bm{Q}_\nu = M_{\nu}$ and $1 \leq \nu \leq 3$)
four-sublattice chiral AFM ordering is favored relative to a single-$\bm{Q}$ ordering because
%the different ordering wave vectors connect \emph{independent} pieces of Fermi surface. In other words, a
%the triple-$\bm{Q}$ ordering 
it simultaneously gaps out the independent pieces of Fermi surface  connected by the symmetry related ordering wave vectors $\bm{Q}_1$, $\bm{Q}_2$ and $\bm{Q}_3$. This argument can in principle be extended to other electron filling fractions that lead to smaller ordering wavectors $Q= |\bm{Q}_{\nu}|$ due to the smaller size of the Fermi surface [see Fig.~\ref{fig:nc}(a)]. Indeed, zero-field SkX with two skyrmions per magnetic unit cell have been reported in TKLMs for smaller filling fractions~\cite{OzawaR2017}. More recently, magnetic field-induced SkX with one skyrmion per magnetic unit cell were reported in the low-density and weak-coupling regimes of the KLM on hexagonal lattices by adding a small easy-axis spin anisotropy~\cite{WangZ2020}.

Based on these observations, we  conjecture that both zero field and field-induced
%triple-$\bm{Q}$ orderings with smaller
%$Q= |\bm{Q}_{\nu}|$ values [see Fig.~\ref{fig:nc}(a)], implying that
%and the chiral version of those orderings corresponds to a triangular skyrmion crystal, we can
SkX with longer  lattice constants (or wavelength) should be rather ubiquitous ground states of  the \emph{isotropic} TKLM in the \emph{intermediate-coupling} regime for a continuous range of electron filling fractions~\cite{BatistaCD2016_review}. A key observation is that the effective 4-spin interactions (or  more generally $2n$-spin interactions with $n>1$) that are generated in this regime are expected to produce an ``attraction'' between Fourier components with different ordering wave vectors $\bm{Q}_{\mu}$ and $\bm{Q}_{\nu}$ with $\mu \neq \nu$. The attractive nature of the effective interaction between different modes reflects the energy gain associated with simultaneously gapping out independent pieces of the Fermi surface. 

While our conjecture is supported by an increasing number of experimental results on centrosymmetric metallic magnets~\cite{YuX2012,YuXZ2014,MallikR1998a,SahaSR1999,KurumajiT2019,ChandragiriV2016,HirschbergerM2019,KhanhND2020,AmorosoD2020,ChakrabarttyD2022,ShangT2021,KanekoK2021,TakagiR2022}, numerical studies of the KLM have only found SkX for fine-tuned sets of Hamiltonian parameters~\cite{OzawaR2017}.
%For a long time, this discrepancy challenges the validity of explaining SkX formation using KLM with generic model parameters.
According to the results presented in this work,   this  requirement arises from finite size effects 
%of fine tuning Hamiltonian parameters is due to the challenge of 
that limit the accuracy of numerical results for the general case.
%obtaining accurate enough numerical results for the general case.
%more general values of the Hamiltonian parameters and filling fractions.
Because of this difficulty, recent theoretical studies of SkX formation have focused on phenomenological models that assume a certain form and sign of 4-spin interactions~\cite{HayamiS2021_review}, which are expected to mimic the effective spin-spin interactions generated by the KLM. Since a perturbative treatment of the Kondo exchange  breaks down beyond second order at low enough temperatures~\cite{AkagiY2012,HayamiS2017},  higher order spin interactions are normally incorporated in an ad-hoc manner~\cite{HayamiS2021_review} or by fitting first principles calculations~\cite{Kurz01,HeinzeS2011}.
Alternatively, the KLM has also been studied in the double exchange limit~\cite{KumarS2010, RejaS2015}, where additional spin-orbit couplings were introduced to facilitate the stabilization of SkX~\cite{KathyatDS2020, KathyatDS2021}.

Only very recently, it was demonstrated that magnetic field induced SkX can naturally emerge in the weak-coupling limit of the KLM for generic filling fractions and model parameters if an easy-axis spin anisotropy~\cite{WangZ2020} or thermal fluctuations~\cite{MitsumotoK2022} are present.
In this limit, the  KLM can be reduced to the  RKKY model~\cite{RudermanMA1954,KasuyaT1956,YosidaK1957}. Away from this regime,  effective  $2n$-spin interactions with $n>1$ become relevant, and the above-mentioned arguments suggest that if the ordering wave vectors connect independent pieces of the Fermi surface, these multi-spin interactions should favor the formation of multi-${\bm Q}$ magnetic orderings. 

%For instance, 4-spin interactions are expected to produce an effective attraction between pairs of modes with the different (symmetry related) ordering wave-vectors. It is then natural to ask if these effective 4-spin interactions can stabilize SkX even in absence of easy-axis anisotropy.

%it remains to be demonstrated that these multi-spin interactions can stabilize SkX even in absence of easy-axis anisotropy.

%Identifying the ingredients that stabilize SkX
%is a main challenge  because new stabilization mechanisms can be accompanied by novel  properties. For instance, the vector chirality is fixed in the magnetic skyrmions of chiral magnets,  while it is a degree of freedom in the SkX of centrosymmetric materials, such as BaFe$_{1-x-0.05}$Sc$_x$Mg$_{0.05}$O$_{19}$, La$_{2-2x}$Sr$_{1+2x}$Mn$_2$O$_7$, Gd$_2$PdSi$_3$ and Gd$_3$Ru$_4$Al$_{12}$~\cite{YuX2012,YuXZ2014,MallikR1998a,SahaSR1999,KurumajiT2019,ChandragiriV2016,HirschbergerM2019}. In the former case, the underlying spiral structure
%arises from competing ferromagnetic and  Dzyaloshinskii-Moriya (DM) interactions~\cite{Dzyaloshinsky1958,Moriya1960}. In contrast, the spiral ordering of centrosymmetric materials arises from  competing  isotropic exchange couplings or dipolar interactions~\citep{OkuboT2012,Leonov2015,LinSZ2016,Hayami16,BatistaCD2016_review}.

The numerical challenge of obtaining a $T=0$ phase diagram of the KLM arises from the smallness of the \emph{effective}  interactions between localized moments in comparison with the \emph{bare} Hamiltonian parameters. Even the small size effects associated with relatively large finite lattices can alter the relative stability of two competing orders in comparison to the thermodynamic limit.
%significant finite-size effects associated with the smallness of the effective interactions between localized moments.
This situation persists  for a Kondo exchange interaction $J$  comparable to the nearest-neighbor hopping $t$,
%the effective interactions can be several orders of magnitude smaller than the bare interactions,
posing a serious challenge for numerical techniques that are implemented on finite lattices.  Here we avoid these undesirable size effects by implementing a novel variational method {\it in the thermodynamic limit}.
This method reveals that SkX are indeed ubiquitous ground states of the \emph{isotropic} TKLM induced by relatively large multi-spin interactions  caused by the above-mentioned Fermi surface effects~\cite{BatistaCD2016_review}.
The field induced SkX phases emerge above a critical coupling strength $J/t$, indicating that moving away from the RKKY regime of $f$-electron magnets should favor the stabilization these topological spin textures relative to other magnetic orderings.

\section{Model and Results}

We consider a 2D TKLM with  \emph{classical} local magnetic moments $\bm{S}_{i}$:
\begin{equation}
\mathcal{H}= -t \sum_{\langle i j\rangle,\sigma} \left(c_{i\sigma}^{\dagger}c_{j\sigma}+h.c.\right) +J\sum_{i,\alpha\beta }c_{i\alpha}^{\dagger}\bm{\sigma}_{\alpha\beta}c_{i\beta}\cdot\bm{S}_{i}
-H\sum_{i}S_{i}^{z}+D\sum_{i}\left(S_{i}^{z}\right)^{2},
\label{eq:Kondo}
\end{equation}
%- \mu \sum_{i \sigma} c_{i \sigma}^{\dagger}c_{i \sigma}
where the operator $c_{i\sigma}^{\dagger}$ ($c_{i\sigma}$) creates (annihilates)
an itinerant electron on site $i$ with spin $\sigma$, and $t>0$ is the hopping amplitude of the nearest neighbor bonds.
%and $t$ is the nearest neighbor hopping amplitude.
%and $c_{\bm{k}\sigma}^{\dagger}$ ($c_{\bm{k}\sigma}$) is the corresponding Fourier transform.
%$\epsilon_{\bm{k}}$ is the bare electron dispersion with chemical potential $\mu$.
The Kondo exchange $J$ couples the local magnetic moments $\bm{S}_{i}$
to the conduction electrons ($\bm{\sigma}$ is the vector of the
Pauli matrices). The last two terms represent a Zeeman coupling to an external field $H$
%$h$ ($H = g \mu_B h$)
and an easy-axis single-ion anisotropy ($D\le 0$).
For the centrosymmetric rare-earth based SkX, the moments of the localized spin (Gd$^{3+}$, Eu$^{2+}$) are quite large ($J=7/2$), which greatly suppresses both the Kondo screening and the quantum fluctuations. With these cases in mind, a classical treatment of the local moments $\bm{S}_i$ is used throughout this paper, and we use the normalization condition $\left|\bm{S}_{i}\right|=1$.

As we mentioned earlier, the goal of this paper is to study the SkX formation in centrosymmetric metallic magnets for which the KLM provides an appropriate description of the low-energy physics. An important difference between these systems and magnetic Mott insulators deep inside the Mott regime is that effective models of  the latter  are dominated by short-ranged two-spin  interactions, and the stabilization of  SkX requires the combination of  magnetic frustration and thermal fluctuations or magnetic anisotropy~\cite{OkuboT2012,LeonovAO2015,LinSZ2016}.

%{\it Method.}
%To prepare for the variational calculation of the $T=0$ phase diagrams,
We start by
%fine tune the filling fraction $n_c$ in order to produce optimal ordering wave vectors with a specific wave number $Q\equiv |\bm{Q}_{\nu}|$.
computing the change of the ordering wave vectors $\bm{Q}_{\nu}$ as a function of the electron filling fraction $n_c$.
In the weak-coupling limit ($J \ll t$), the effective spin-spin interactions are described by an RKKY Hamiltonian and the ordering
wave vectors are obtained by maximizing the Lindhard function~\cite{WangZ2020}. However, a different method must be used to find the values of $\bm{Q}_{\nu}$ away from this regime because
higher order spin interactions produce a significant renormalization of the
ordering wave vectors $\bm{Q}_{\nu}$~\footnote{See Ref.~\cite{WangZ2016a} Fig.~(S2) for the dependence of $Q$ on $J/t$ at a fixed filling fraction.}.
%These contributions renormalize the ordering vector
%Due to the renormalization of the ordering wave vectors from the finite Kondo coupling, the simple approach of analyzing the  no longer provides a good estimate.

\begin{figure}
\centering
\includegraphics[width=0.7\columnwidth]{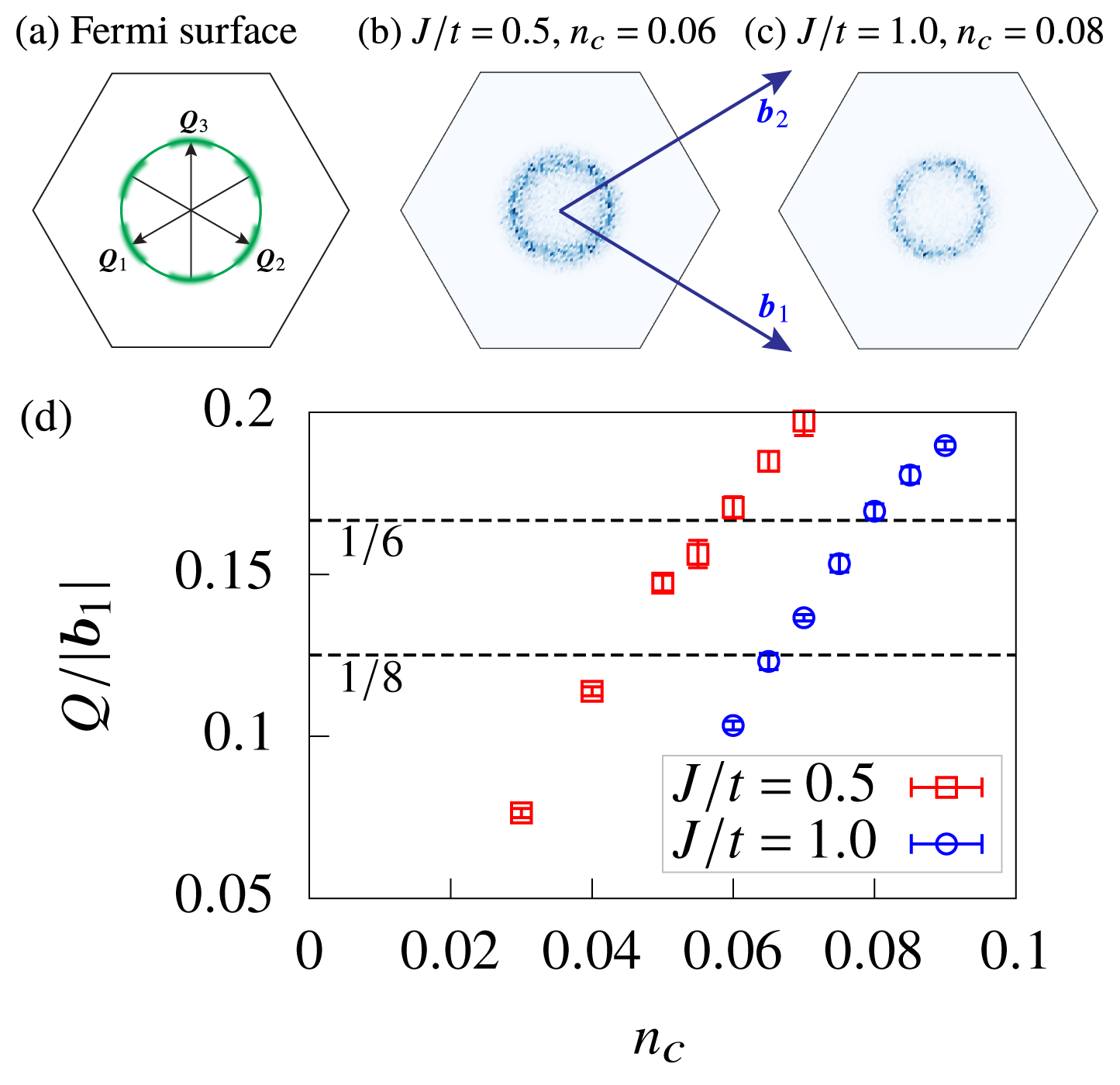}
\caption{(a) Illustration of independent pieces of Fermi surface that are connected by the symmetry related ordering wave vectors \{$\bm{Q}_1$, $\bm{Q}_2$,  $\bm{Q}_3$\} at low filling fractions. (b)-(d) KPM-SLL results for the TKLM on a $96\times96$ lattice with $H=D=0$ and $T=10^{-5} J^2/t$. (b)-(c) Snapshots of the static spin structure factor $\mathcal{S}(\bm{q})$. \{$\bm{b}_1$, $\bm{b}_2$\}  are the basis vectors in the reciprocal space. (d) Ordering wave number as a function of the filling fraction. }
\label{fig:nc}
\end{figure}

%A method that is able 
The values of $\bm{Q}_\nu$ can be computed in an unbiased manner by iterating two steps at a given temperature. One of the steps  ``integrates out'' the conduction electrons by evaluating the free energy of  a proposed spin configuration. The other  step consists of a stochastic update the spin configuration based  on the value of the free energy. The simplest implementation of these two steps  combines  exact diagonalization (ED) to compute the free energy and a Metropolis Monte Carlo algorithm to update the spin configurations. However, this combination can only be applied to small lattices because of the $\mathcal{O}(N^3)$ cost of ED~\cite{YunokiS1998}. The kernel polynomial method (KPM)~\cite{WeisseA2006_RMP} replaces the ED step with an approximated polynomial expansion of the single-electron density of states for the given spin configuration.
%evaluates the physical quantities through orthogonal polynomial expansion instead of direct diagonalization, 
%that is able to reduce the $\mathcal{O}(N^3)$ scaling of ED to $\mathcal{O}(N)$, and 
Because the numerical cost of KPM scales linearly with $N$, the KPM method  has been successfully applied to simulations of  KLMs~\cite{MotomeY1999,MotomeY2000}. Another numerical improvement based on the KPM method is 
a stochastic evaluation the local torques (gradients of the free energy) via automatic differentiation, enabling an implementation of Langevin dynamics that greatly speeds up the simulations~\cite{BarrosK2013}. The efficiency of the KPM can be further improved by employing a gradient-based probing method~\cite{TangJM2012,WangZ2018}, that enabled us to simulate KLMs with $\mathcal{O}(10^4)$ lattice sites.

Here we employ a variant of the KPM that utilizes gradient-based probing~\cite{BarrosK2013,WangZ2018} to obtain an unbiased estimate of the wave vector of low-energy spin configurations on  finite %TKLM
lattices of $96\times 96$ sites. 
For $H=D=0$, $T=10^{-5}J^2/t$ and $J/t=\{0.5,\,1.0\}$,  we integrate the dimensionless stochastic Landau-Lifshitz (SLL) dynamics with a unit damping parameter using the Heun-projected scheme for a total of 45000 steps of duration $\Delta \tau=0.5 / (J^2/t)$. The order of the Chebyshev polynomial expansion is $M=1000$ and we use the gradient-based probing method with $S=256$ colors~\cite{WangZ2018}. The first 30000 steps are discarded for equilibration and the rest 15000 steps are used for measurements. At each temperature, we average over 6 independent runs to estimate the error bars. Finally, the values of $\bm{Q}_\nu$ can be estimated from the peak positions of the static spin structure factor, defined as
\begin{equation}
\mathcal{S}(\bm{q}) \equiv \langle \bm{S}_{\bm{q}} \cdot \bm{S}_{-\bm{q}} \rangle,
\end{equation}
where $\bm{S}_{\bm{q}} \equiv \frac{1}{\sqrt{N}} \sum_{i} e^{-\iu \bm{q}\cdot \bm{r}_i} \bm{S}_i$ is the Fourier transform of the real-space spin configurations, and $N$ is the total number of lattice sites.

Figures~\ref{fig:nc}(b)-(c) show  typical snapshots of $\mathcal{S}(\bm{q})$ near the end of the KPM-SLL simulation. For the low filling fractions that we are considering, $\mathcal{S}(\bm{q})$ takes its maximum value on a ring~\cite{WangZ2016a}.
%with wave number $Q$ that depends on the filling $n_c$.
Note that, in equilibrium, spin configurations with broken discrete symmetries are generally allowed at finite low temperatures~\cite{OkuboT2012}. As we will see later,
%in this paper,
the lack of symmetry breaking in our KPM-SLL simulation is due to the small energy scale of the effective interactions between local moments
%difference between competing low energy states
 in units of the bare interactions of the TKLM.
%that leads to a significant finite size effect.
%and failure in finding the converged states in most finite-size simulations of the KLM.

%On the other hand,
The magnetic unit cell for each filling fraction can be inferred from the  $Q (n_c)$ curve produced by the KPM-SLL simulation,  allowing us to %fix the magnetic unit cell and
exploit  the translational symmetry of the optimal spin configuration and find the ground state in the thermodynamic limit. As indicated by the dashed lines of Fig.~\ref{fig:nc}(d), we can always find commensurate ordering wave vectors $\bm{Q}_{\nu} (n_c)$
%that are commensurate with the lattice period
by choosing the right filling fraction. For example, for $Q=|\bm{b}_1|/L$, the magnetic unit cell contains $L\times L$ spins spanned by the basis \{$L\bm{a}_1$, $L\bm{a}_2$\}, where $\bm{a}_1$ and $\bm{a}_2$ are primitive vectors of the triangular lattice~\footnote{For $J/t\rightarrow 0$, this becomes identical to the susceptibility analysis in Ref.~\cite{WangZ2020}.}.

The $T=0$ energy density $e$ of each periodic spin configuration is computed by diagonalizing $\mathcal{H}$ in momentum space and integrating the  sum  of energies of occupied single-particle states over the reduced Brillouin zone $\mathcal{B}_r$:
\begin{equation}
e= \frac{1}{L^2} \sum_{n=1}^{2L^2} \int_{\mathcal{B}_r} \frac{d\tilde{\bm{k}}}{\mathcal{A}_{\mathcal{B}_r}} \Theta\left[\mu-\epsilon_{n}(\tilde{\bm{k}})\right]\epsilon_{n}(\tilde{\bm{k}})
 +\frac{1}{L^2} \sum_{\bm{R}} \left[-H S_{\bm{R}}^{z}+D \left(S_{\bm{R}}^{z}\right)^{2}\right], \label{eq:energy_integral}
\end{equation}
where $\Theta(x)$ is the Heaviside step function, and ${\mathcal{A}}_{{\mathcal{B}}_r}$ is the area of $\mathcal{B}_r$.
%and $N_b=2 L^2$ is the number of bands and $L^2$ is the number of spins in the magnetic unit cell.

For convenience, we mainly focus on magnetic orderings with $L=6$~\footnote{The choice of commensurate values of $Q$ is only for the convenience of calculation (nearby incommensurate values of $Q$ should produce similar phase diagrams).}, which correspond to $n_c\approx 0.0586$ for $J/t=0.5$ and $n_c \approx 0.0796$ for $J/t=1$ according to Fig.~\ref{fig:nc}. For any fixed parameter set \{$H$, $D$\}, we minimize the energy density $e$ with respect to $2 L^2$ variational parameters (each $\bm{S}_{\bm{R}}$ is parametrized by two independent angles) at fixed filling fraction $n_c$. To locate the global minimum, we perform many independent runs with different random initial spin configurations (typically 20) and keep only the lowest energy solution (see Appendix~\ref{sec:variational} for details).

%The magnitude of the ordering wave vector is $Q = l |\bm{b}_1|/L$, where $l$ is the minimum integer that satisfies the equation.
%All possible spin configurations are parametrized by  that are determined by finding the global minimum of the energy density $e$~\cite{nlopt,Nocedal1980,Liu1989}.

\begin{figure}
\centering
\includegraphics[width=\columnwidth]{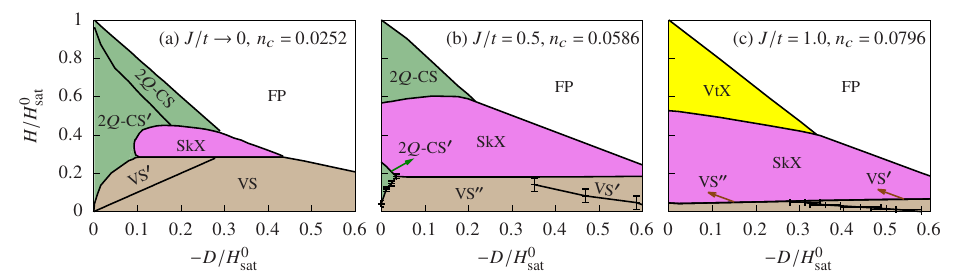}
\caption{Phase diagrams of the TKLM with easy-axis single-ion anisotropy
in a magnetic field with the ordering wave number $Q=|\bm{b}_1|/6$, where the RKKY limit (a) is taken from Ref.~\cite{WangZ2020}. The error bars of phase boundaries at low field indicate the limited numerical accuracy due to quasi-degenerate states. The saturation field at $D=0$ is $H_\text{sat}^0=8.53 \times 10^{-4}t$ for (b) and $H_\text{sat}^0=4.31 \times 10^{-3}t$ for (c).
\label{fig:phd}}
\end{figure}

The results of the variational calculation for $J/t=\{0.5,\,1\}$ are summarized in Fig.~\ref{fig:phd}. In comparison  to the
weak-coupling regime  ($J/t \ll 1$) described by an effective RKKY Hamiltonian [Fig.~2(b) of Ref.~\cite{WangZ2020}, reproduced here as Fig.~\ref{fig:phd}(a)], it is clear that the SkX phase expands upon increasing the Kondo coupling $|J/t|$. This phenomenon highlights the crucial role played by higher order spin interactions~\cite{BatistaCD2016_review,HayamiS2021_review}. An important consequence of the strengthening of these interactions relative to two-spin interactions  is that the single-ion anisotropy is no longer necessary to stabilize the SkX. We have also confirmed that the SkX remains  stable for $J/t=1$ and $D=0$ when the filling fraction is reduced to obtain a smaller  ordering wave vector $Q=|\bm{b}_1|/8$ (longer wave length).

In the RKKY limit, we found that the optimal wavevector remains practically unchanged for finite $D$ and $H$, indicating that the variational calculation with fixed $L$ remains accurate over the full phase diagram~\cite{WangZ2020}. To verify if this is still true for finite $J/t$, we consider a point inside the $J/t=0.5$ phase diagram [Fig.~\ref{fig:phd}(b)]: $H=-2D=0.469H_{\text{sat}}^0$. The KPM-SLL simulations indicate that a slightly different filling fraction $n_c\approx 0.0548$ (the deviation is comparable to the errorbar of the KPM-SLL simulation) is required  to keep the ordering wave vector $Q/|\bm{b}_1|=1/6$ unchanged.
%slightly deviated from the $H=D=0$ value .
The variational calculation for the new filling fraction $n_c\approx 0.0548$ at $H=-2D=0.469H_{\text{sat}}^0$ confirms that the SkX is still the ground state.

\begin{figure}
\centering
\includegraphics[width=0.7\columnwidth]{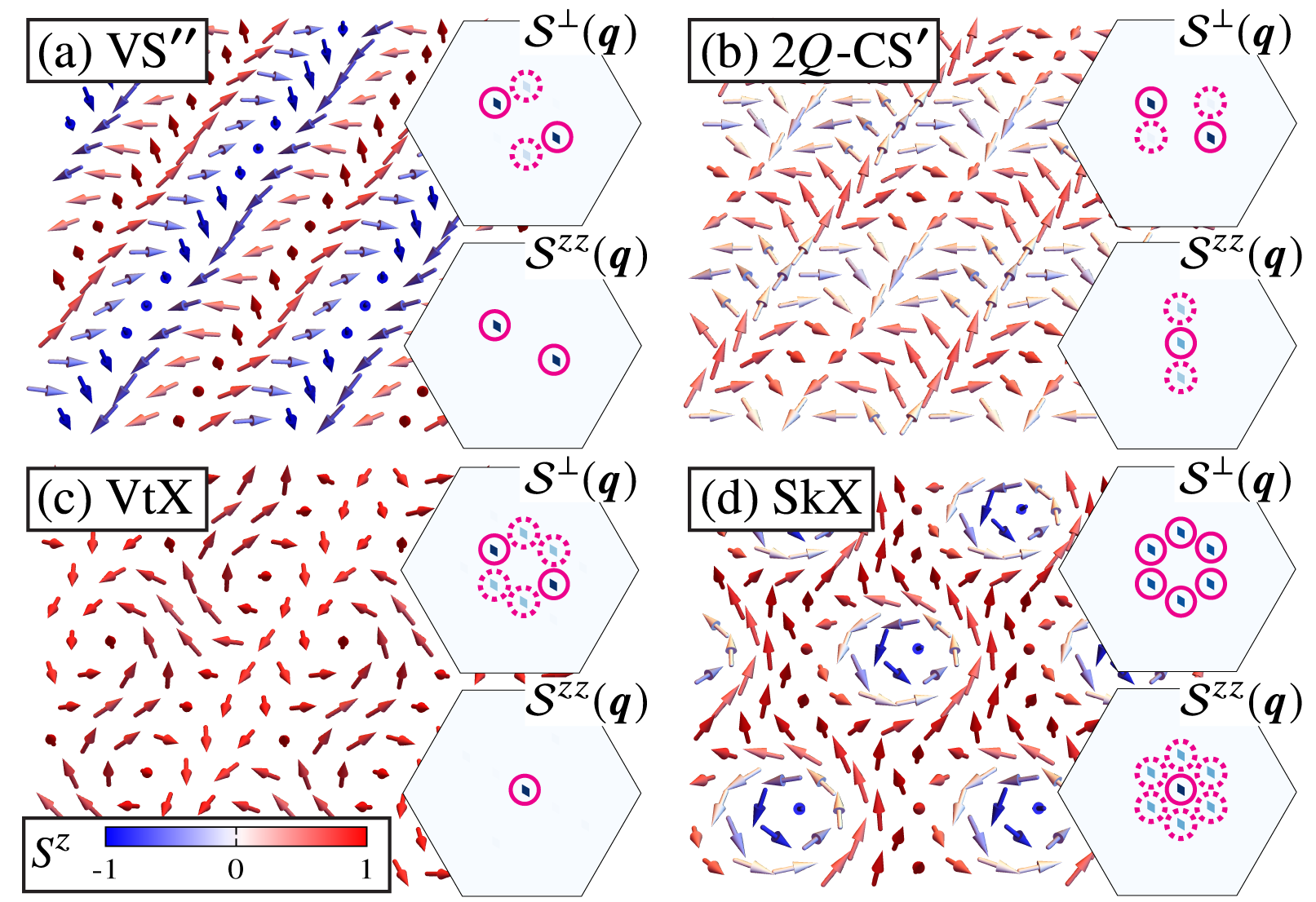}
\caption{Spin configurations of phases shown in Fig.~\ref{fig:phd}. The insets show the in-plane ($\mathcal{S}^{\perp}$)
and out-of-plane ($\mathcal{S}^{zz}$) static structure factors in
the first Brillouin zone. The solid (dotted) circles highlight the dominant (subdominant) peaks. The spin configurations of VS$^\prime$ and $2Q$-CS can be found in Ref.~\cite{WangZ2020}.
\label{fig:spins}}
\end{figure}

Besides further stabilizing the SkX phase, the larger Kondo exchange also produces a few multi-$\bm{Q}$ phases that do not appear in the RKKY limit~\cite{WangZ2020}. In particular, a new vertical spiral phase (VS$^{\prime \prime}$) [see Fig.~\ref{fig:spins}(a)] and a vortex crystal phase (VtX) [see Fig.~\ref{fig:spins}(c)] appear at low and high fields respectively (see Appendix~\ref{sec:fourier} for the Fourier analysis of the VS$^{\prime \prime}$ and VtX states). We note that the distribution of local scalar spin chirality, $\bm{S}_{i} \cdot (\bm{S}_j \times \bm{S}_k)$, on each triangular plaquette exhibits chiral stripes with alternating signs (zero net scalar chirality) in the VS$^{\prime \prime}$, $2Q$-CS and $2Q$-CS$^\prime$. Similar stripes of scalar chirality have been reported in Refs.~\cite{SolenovD2012,OzawaR2016}.
Vortex crystals below the saturation field have also been reported for spin models with short-range anisotropic exchange interactions~\cite{KamiyaY2014,WangZ2015,WangZ2021}.

To date, there are only two  direct
numerical confirmations of the SkX ground states in the TKLM~\eqref{eq:Kondo}. The first one is the four-sublattice  chiral ordering~\cite{MartinI2008,AkagiY2010,KatoY2010,BarrosK2013} which is strictly speaking the short wavelength limit ($L=2$ and $Q=|\bm{b}_1|/2$) of the SkX on a triangular lattice. The  second one is directly obtained from the KPM-SLL method by fine-tuning  both the third neighbor hopping and the  chemical potential~\cite{OzawaR2017}.
Short wavelength multi-$\bm{Q}$ chiral spin structures on kagome KLM were also reported with fine-tuned Fermi surfaces~\cite{BarrosK2014}.
The fine-tuning is required to get sharp peaks of the Lindhard function and maximize the magnitude of the effective spin-spin interactions. It is then natural to ask why previous numerical attempts were not able to identify the ubiquitous nature
of triple-$\bm{Q}$ SkX orderings in the TKLM.
%To this point, it is natural to ask where the previous numerical difficulty comes from, and why the variational method in this paper is able to determine the low energy states for general parameters.

\begin{figure}
\centering
\includegraphics[width=0.95\columnwidth]{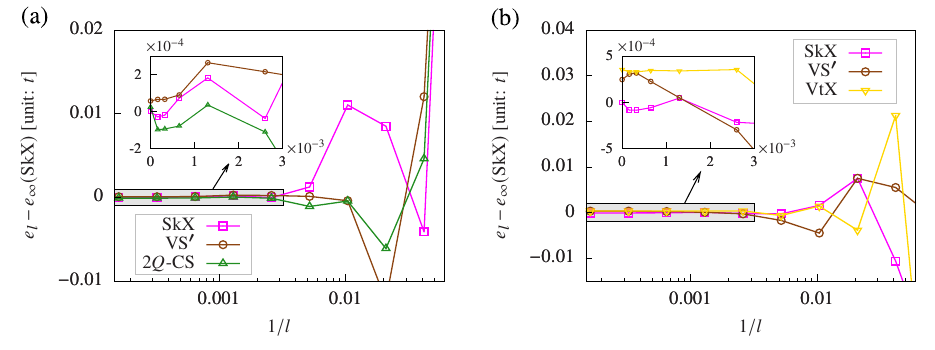}
\caption{The energy densities $e_l$ of different states evaluated on a uniform $(l/L)^2$ mesh in the reduced Brillouin zone $\mathcal{B}_r$, shifted by the energy density of the SkX state $e_\infty$(SkX) evaluated by the integral Eq.~\eqref{eq:energy_integral}.
The data points at $1/l=0$ are obtained from numerical integration instead of discrete sum.
In (a), the three states are obtained by the variational method for the same parameter set \{$J/t=0.5$, $n_c=0.0586$, $D/t=-10^{-4}$, $H/t=3.125\times 10^{-4}$, $L=6$\}, converged from different random initial spin configurations; Similarly in (b), the three states are obtained by the variational method for the same parameter set \{$J/t=1.0$, $n_c=0.0796$, $D/t=-10^{-3}$, $H/t=10^{-3}$, $L=6$\}, converged from different random initial spin configurations. For each parameter set, the SkX state is the ground state in the thermodynamic limit, while the other two states are metastable (local energy minimum).
}
\label{fig:finite}
\end{figure}

The key observation is that the effective spin interactions (RKKY and higher order terms) are much  smaller than the bare coupling constant $J$, as it is clear from the values of the saturation fields in Fig.~\ref{fig:phd}~[$H_\text{sat}^0=8.53 \times 10^{-4}t$ for (b) and $H_\text{sat}^0=4.31 \times 10^{-3}t$ for (c)].
%is a low energy description of the original KLM.
Furthermore, the  splitting between competing single and multi-$\bm{Q}$ orderings is controlled by an even smaller energy scale that results from the competition between effective higher order spin interactions and the small RKKY energy cost of  higher harmonic components of generic multi-$\bm{Q}$ orderings, which are required  to fulfill the normalization constraint $|\bm{S}_i|=1$.  This situation leads to  very small    differences of the energy density, $\Delta e$, in  units of the hopping amplitude $t$. Such small energy differences are sensitive to numerical accuracy, which is often limited by system size and by other approximations of the numerical method (e.g., the order of the Chebyshev polynomial expansion and number of random vectors in the KPM-SLL method). To estimate the error introduced by the finite size effects, we simply need to replace the integral in Eq.~\eqref{eq:energy_integral} with a discrete sum on a uniform $(l/L)^2$ grid in $\mathcal{B}_r$. The discrete sum corresponds to the energy density  of the given spin configuration on a finite lattice of $N=l^2$ sites. In Fig.~\ref{fig:finite}(a), we consider three converged spin configurations, corresponding to a stable SkX and metastable  VS$^\prime$ and 2$\bm{Q}$-CS solutions obtained with the variational method for the same parameter set and different random initial spin configurations. The  dependence of the energy densities on the linear lattice size, $e_l$,  clearly indicates that for $J/t=0.5$ it is necessary to consider finite lattices with $l \gg 6000$ to obtain results that are representative of the thermodynamic limit. Even For $J/t=1$, Fig.~\ref{fig:finite}(b) shows that a lattice with $l \gg 1000$ is required to achieve convergence, which is beyond the typical system size that can be reached with the KPM-SLL method.
%which explains the failure of the finite size calculations on the Kondo Lattice models.
In other words, it is essential to take the thermodynamic limit in Eq.~\eqref{eq:energy_integral} (performing high-accuracy integration) in order to find  the correct low energy states of the KLM for general sets of parameters.

\section{Conclusion}
A key outcome of this work is the confirmation that effective 4-spin and higher order interactions generated in KLMs with hexagonal symmetry lead to SkX  that span a large spectrum of wavelengths ranging  from $\lambda \gg a$ for small Fermi surfaces~\cite{WangZ2020} to the short wave length or dense limit $\lambda =2a$ corresponding to the ``tetrahedral'' ordering reported in Refs.~\cite{MartinI2008,AkagiY2010,KatoY2010}. Based on our results, we conjecture that SkX of intermediate wave lengths between six and two lattice parameters (e.g. $\lambda= 4a$) should also emerge above a critical value of $J/t$ for some range of filling fractions between $n=0.06$ and $n=0.25$. We also expect SkX to emerge for even higher filling fractions with more complicated Fermi surfaces. In other words, SkX can be ground states without the need of fine tuning  the Fermi surface to produce sharp peaks in the spin susceptibility of the conduction bands.
%We emphasize that previous numerical works~\cite{KatoY2010,BarrosK2013,BarrosK2014,OzawaR2017} were restricted to this special case because of limitations imposed by the above-mentioned size effects.
The novel approach that is introduced here avoids this restriction because the results are directly obtained in the thermodynamic limit. Unlike other approaches in which the magnetic unit cell is fixed in an ad-hoc fashion~\cite{AkagiY2010}, we have used unbiased numerical simulations to determine the optimal magnetic unit cell for a given filling fraction of the conduction bands.

It is worth noting that SkX with both topological charge $n_\text{sk}=1$ and $n_\text{sk}=2$ per magnetic unit cell were reported before, and the $n_\text{sk}=2$ case is often stabilized at zero field~\cite{MartinI2008,OzawaR2017}. Since the stabilization of both kinds of phases occurs via the generation of 4-spin interactions that favor triple-$\bm{Q}$ magnetic orderings, our results suggest that SkX of both types should generally emerge for a sequence of electron filling fractions $n_c$ that connect the relatively long ($n_c \ll 1$) and short wave length ($n_c \sim 1$) regimes.

%We have recently demonstrated~\cite{WangZ2020} that SkX can naturally emerge in weakly-coupled hexagonal \emph{easy-axis} $f$-electron magnets, for which the Kondo Lattice model (KLM) can be reduced to a Ruderman-Kittel-Kasuya-Yosida (RKKY) model~\citep{RudermanMA1954,KasuyaT1956,YosidaK1957}.
%The recent discovery of rare earth based materials Gd$_2$PdSi$_3$ and Gd$_3$Ru$_4$Al$_{12}$ that contain a magnetic field induced SkX phase in their phase diagrams~\cite{MallikR1998a,SahaSR1999,KurumajiT2019,ChandragiriV2016,HirschbergerM2019} motivates us to further explore  the KLM away from the RKKY limit.

%In the adiabatic limit, the momentum space Berry curvature is controlled by a real space Berry curvature that is proportional to the skyrmion density: each skyrmion produces an effective flux equal to the flux quantum $\Phi_0$. Consequently, Hall conductivities comparable to the quantized value ($e^2/h$)  can in principle be achieved if the ordering wave vector of the SkX is comparable to the Fermi wave vector $k_F$.

Establishing  a \emph{generic} stabilization mechanism of SkX in KLMs is very important because  most of the known magnetic SkX have been reported in metals~\cite{MallikR1998a,SahaSR1999,YuX2012,YuXZ2014,ChandragiriV2016,KurumajiT2019,HirschbergerM2019}, where the exchange interaction between  magnetic moments and conduction electrons enables novel responses, such as the topological Hall effect (THE)~\cite{OnodaM2004,YiSD2009,HamamotoK2015,GobelB2017} and current-induced skyrmion motion~\cite{JonietzF2010,YuXZ2012,SchulzT2012,NagaosaN2013}. The THE arises from the Berry curvature acquired by the reconstructed electronic bands. A crucial distinctive character of the SkX that emerge in KLMs is that topological Hall effect can be extremely large (comparable to the quantized value) because their lattice spacing is dictated by the Fermi wave vector~\cite{KurumajiT2019}.

Demonstrating the generic nature of the mechanism  is particularly relevant because
real materials  comprise multiple conduction bands and more general forms of Kondo interaction~\footnote{While the variational method introduced in this paper can be straightforwardly applied to more realistic cases, the enlarged matrix size due to multiple conduction bands poses an additional numerical challenge, which can be alleviated by using  modern parallel computing hardware.}.
The key observation is that
four and higher-order effective spin interactions  generated by processes that involve {\it independent} pieces of the Fermi surface  connected by symmetry related wave vectors favor multi-$\bm{Q}$ spin configurations [see Fig.~\ref{fig:nc}(a)]. While the amplitude of these effective higher order spin interactions cannot be obtained from perturbation theory
because they are non-analytic functions of $J/t$
~\cite{AkagiY2012,BatistaCD2016_review,OzawaR2017,HayamiS2017}, they can in principle be calculated using other techniques, such as resummation of diagrammatic series.
Since their strength  relative to the  two-spin  interaction grows with $J/t$, we expect that, \emph{in absence of easy-axis single-ion anisotropy}, field induced SkX and other multi-$\bm{Q}$ orderings should emerge above a critical value of $J/t$. This conclusion is supported by an increasing number of experimental results in $f$-electron magnets~\cite{XuY2021,SeoS2021,ZhangH2021a,MoyaJM2022}.
We note that $d$-electron systems with localized moments in $t_{2g}$ orbitals coupled via Hund's exchange to conduction $e_g$ electrons can also provide natural realizations of the intermediate-coupling regime $J \gtrsim t$ considered in this work~\cite{WangW2016,LiH2019,WangS2020,ZhengX2021,DallyRL2021,NabiMRU2021}.

% LiH2019 (Fe3Sn2: seems not centrosymmetric by eye. However some paper say it's centrosymmetric while some say not... Confused...)

\section*{Acknowledgements}
We thank Shi-Zeng Lin and Kipton Barros for helpful discussions.
ZW was supported by funding from the Lincoln Chair of Excellence in Physics.
During the writing of this paper, ZW was supported by the U.S. Department of Energy through the University of Minnesota Center for Quantum Materials, under Award No.~DE-SC-0016371. CDB acknowledges support from U.S. Department of Energy, Office of Science, Office of Basic Energy Sciences, under Award No.~DE-SC0022311.
This research used resources of the Oak Ridge Leadership Computing
Facility at the Oak Ridge National Laboratory, which is supported
by the Office of Science of the U.S. Department of Energy under Contract No.~DE-AC05-00OR22725.

\begin{appendix}

\section{Variational Method}
\label{sec:variational}
In this section we consider the $T=0$ commensurate states. Furthermore, we assume that the magnetic unit cell is spanned by the basis \{$L\bm{a}_1$, $L\bm{a}_2$\}, where $\bm{a}_1$ and $\bm{a}_2$ are the primitive vectors of the triangular lattice. Note that the value of $L$ can be estimated from unbiased techniques, such as the KPM-SLL method~\cite{BarrosK2013,WangZ2018} used in this paper, which can be implemented  in relatively large lattices.

In the following, we will label different sublattices of the magnetic unitcell by $\bm{R}$, and different unitcells by $\tilde{\bm{r}}$, so the coordinates of each site can be expressed as the sum $\bm{r}=\tilde{\bm{r}}+\bm{R}$~\cite{SenechalD2008}. The translational symmetry of these commensurate states allows us to perform the Fourier transform:
\begin{equation}
c_{\tilde{\bm{r}},\bm{R},\sigma}	=\sqrt{\frac{L^2}{N}}\sum_{\tilde{\bm{k}}}e^{\iu\tilde{\bm{k}}\cdot\tilde{\bm{r}}}c_{\tilde{\bm{k}},\bm{R},\sigma},\label{eq:fourier}
\end{equation}
where $N$ is the total number of the lattice sites, and $\tilde{\bm{k}}$ labels the allowed momenta in the reduced Brillouin zone $\mathcal{B}_r$.

%It is clear from Eq.~\eqref{eq:fourier} that the fermionic operators in different unit cells are connected by the Bloch's theorem:
%\begin{equation}
%c_{\tilde{\bm{k}},\bm{R}+\tilde{\bm{r}}^{\prime},\sigma}=e^{\iu\tilde{\bm{k}}\cdot\tilde{\bm{r}}^{\prime}}c_{\tilde{\bm{k}},\bm{R},\sigma},
%\end{equation}
%which is crucial to evaluate terms that involve hopping between different magnetic unitcells.

The KLM considered in the main text becomes block-diagonal in  $\tilde{\bm{k}}$-space, $\mathcal{H} = \sum_{\tilde{\bm{k}}} \mathcal{H}_{\tilde{\bm{k}}}$, with
\begin{equation}
\mathcal{H}_{\tilde{\bm{k}}} =
\sum_{\bm{R}} \left[ -t \sum_{\eta}\sum_{\sigma} c_{\tilde{\bm{k}},\bm{R},\sigma}^{\dagger}c_{\tilde{\bm{k}},\bm{R}+\bm{r}_{\eta},\sigma}
+ J \sum_{\alpha\beta}c_{\tilde{\bm{k}},\bm{R},\alpha}^{\dagger}\bm{\sigma}_{\alpha\beta}c_{\tilde{\bm{k}},\bm{R},\beta}\cdot\bm{S}_{\bm{R}}
-H S_{\bm{R}}^{z}+D \left(S_{\bm{R}}^{z}\right)^{2} \right]. \label{eq:model_variational}
\end{equation}
The single-particle eigenstates are obtained by   diagonalizing the  $2L^2 \times 2L^2$ block matrix of the operator $\mathcal{H}_{\tilde{\bm{k}}}$ whose    eigenvalues  are denoted by $\epsilon_{\tilde{\bm{k}},n}$.

The energy density at $T=0$ can be written as
\begin{equation}
e= \frac{1}{N}\sum_{\tilde{\bm{k}}}\sum_{n=1}^{2L^2} \Theta\left(\mu-\epsilon_{\tilde{\bm{k}},n}\right)\epsilon_{\tilde{\bm{k}},n}  +\frac{1}{L^2} \sum_{\bm{R}} \left[-H S_{\bm{R}}^{z}+D \left(S_{\bm{R}}^{z}\right)^{2}\right]. \label{eq:energy_general}
\end{equation}

While Eq.~\eqref{eq:energy_general} is valid for any  system size, it is  crucial to take the thermodynamic limit $N\rightarrow \infty$ to identify  the correct ground state of the  KLM. This is done by converting the discrete sum $\frac{1}{N}\sum_{\tilde{\bm{k}}}$ into the integral:
\begin{equation}
e = \frac{1}{L^2} \int_{\mathcal{B}_r} \frac{d\tilde{\bm{k}}}{\mathcal{A}_{\mathcal{B}_r}} \sum_{n=1}^{2L^2} \Theta\left(\mu-\epsilon_{\tilde{\bm{k}},n}\right)\epsilon_{\tilde{\bm{k}},n} +\frac{1}{L^2} \sum_{\bm{R}} \left[-H S_{\bm{R}}^{z}+D \left(S_{\bm{R}}^{z}\right)^{2}\right],
\label{eq:energy_integral_supp}
\end{equation}
where $\mathcal{A}_{\mathcal{B}_r}$ is the area of the reduced Brillouin zone $\mathcal{B}_r$.
For the canonical ensemble used in this paper, the chemical potential $\mu$ is determined self-consistently at every step of the minimization from  the filling fraction:
\begin{equation}
n_c=\frac{1}{2L^2}\int_{\mathcal{B}_r} \frac{d\tilde{\bm{k}}}{\mathcal{A}_{\mathcal{B}_r}} \sum_{n=1}^{2L^2}\Theta\left(\mu-\epsilon_{\tilde{\bm{k}},n}\right).
\end{equation}

To this point, the meaning of ``variational'' becomes clear. The variational space is defined by $2L^2$ parameters corresponding to the two angles of the classical moment $\bm{S}_{\bm{R}}$. Equipped with a reliable optimization routine~\cite{nlopt}, the only control parameter of accuracy is just $L$. In other words, for any integer $L$, if we can pre-compute the corresponding filling fraction $n_c$ exactly, then the energy expression~\eqref{eq:energy_integral_supp} also becomes exact. Since we pre-computed the relation of $L$ vs $n_c$ from KPM, the deviation from the exact solution is controlled by the accuracy of KPM when we computed the values of $\bm{Q}_\nu$.

The $T=0$ states can be obtained by minimizing $e$ at fixed $n_c$ as a function of these $2L^2$ variational parameters~\cite{nlopt,NocedalJ1980,LiuDC1989}. For local minimization algorithms, the converged results are often metastable local minima (different initial conditions can lead to different final states). For each parameter set, we typically perform 20 independent runs with different random initial spin configurations. 
More runs are required to select out the global minimum near the phase boundaries where the competition between different states is more subtle.

The combination of ED, integration and minimization is computationally expensive. To obtain converged results in reasonable amount of time, it is beneficial to use derivative based minimization algorithms. The derivative of the energy density is given by:
\begin{equation}
\frac{d e}{d\bm{S}_{\bm{R}}} = \frac{1}{L^2} \int_{\mathcal{B}_r}
\frac{d\tilde{\bm{k}}}{\mathcal{A}_{\mathcal{B}_r}} \text{Tr} \left[ f(\tilde{\bm{k}}) \frac{d h(\tilde{\bm{k}})}{d \bm{S}_{\bm{R}}} \right]  + \frac{1}{L^2}\left[-H\begin{pmatrix}0\\
0\\
1
\end{pmatrix}+2D\begin{pmatrix}0\\
0\\
S_{\bm{R}}^{z}
\end{pmatrix}\right],
\end{equation}
where $h(\tilde{\bm{k}})$ is the $2L^2 \times 2L^2$ matrix of the operator $\mathcal{H}_{\tilde{\bm{k}}}$ given in Eq.~\eqref{eq:model_variational},
\begin{equation}
f(\tilde{\bm{k}}) = \sum_{n=1}^{2L^2} \Theta\left(\mu-\epsilon_{\tilde{\bm{k}},n}\right)  |\psi_{\tilde{\bm{k}},n} \rangle \langle\psi_{\tilde{\bm{k}},n} |,
\end{equation}
is the density matrix,
and $|\psi_{\tilde{\bm{k}},n}\rangle$ is the eigenvector with eigenvalue $\epsilon_{\tilde{\bm{k}},n}$.

\section{Fourier analysis}
\label{sec:fourier}
For most of the states that appear in this paper, the Fourier analysis has been documented in Ref.~\cite{WangZ2020}. Here we further analyze the new states that appear in the phase diagram for larger values of $J/t$.

Denote the ordering wave vectors as
\begin{equation}
\bm{Q}_1 = -\bm{b}_2/L,\quad \bm{Q}_2 = \bm{b}_1/L, \quad \bm{Q}_3 = -\bm{Q}_1 - \bm{Q}_2.
\end{equation}

In the VS$^{\prime \prime}$ phase, the normalized spin configurations $\bm{S}_{\bm{r}}=\bm{m}_{\bm{r}} / |\bm{m}_{\bm{r}}|$ can be parametrized as:
\begin{subequations}
\begin{align}
m_{\bm{r}-\bm{r}_{0}}^{x}	&=-a_{1}\cos\phi\sin\left(\bm{Q}_{1}\cdot\bm{r}\right)+a_{2}\sin\phi\sin\left(\bm{Q}_{2}\cdot\bm{r}\right), \\
m_{\bm{r}-\bm{r}_{0}}^{y}	&=-a_{1}\sin\phi\sin\left(\bm{Q}_{1}\cdot\bm{r}\right)-a_{2}\cos\phi\sin\left(\bm{Q}_{2}\cdot\bm{r}\right), \\
m_{\bm{r}-\bm{r}_{0}}^{z}	&=a_{0}-a_{1}\cos\left(\bm{Q}_{1}\cdot\bm{r}\right).
\end{align}
\end{subequations}

The normalized spin configurations of the VtX phase can be parametrized as:
\begin{subequations}
\begin{align}
m_{\bm{r}-\bm{r}_{0}}^{x}	&=a_{1}\sin\phi\sin\left(\bm{Q}_{1}\cdot\bm{r}\right)-a_{2}\cos\phi\left[\cos\left(\bm{Q}_{2}\cdot\bm{r}+\theta\right)-\cos\left(\bm{Q}_{3}\cdot\bm{r}+\theta\right)\right],\\
m_{\bm{r}-\bm{r}_{0}}^{y}	&=a_{1}\cos\phi\sin\left(\bm{Q}_{1}\cdot\bm{r}\right)+a_{2}\sin\phi\left[\cos\left(\bm{Q}_{2}\cdot\bm{r}+\theta\right)-\cos\left(\bm{Q}_{3}\cdot\bm{r}+\theta\right)\right],\\
m_{\bm{r}-\bm{r}_{0}}^{z}	&=a_{0}-a_{3}\sin\left(\bm{Q}_{1}\cdot\bm{r}\right).
\end{align}
\end{subequations}

\end{appendix}

\bibliography{ref}

\begin{thebibliography}{10}
\providecommand{\url}[1]{\texttt{#1}}
\providecommand{\urlprefix}{URL }
\expandafter\ifx\csname urlstyle\endcsname\relax
  \providecommand{\doi}[1]{doi:\discretionary{}{}{}#1}\else
  \providecommand{\doi}{doi:\discretionary{}{}{}\begingroup
  \urlstyle{rm}\Url}\fi
\providecommand{\eprint}[2][]{\url{#2}}

\bibitem{BogdanovAN1989}
A.~N. Bogdanov and D.~A. Yablonskii,
\newblock \emph{Thermodynamically stable ``vortices" in magnetically ordered
  crystals: {{The}} mixed state of magnets},
\newblock Zh. Eksp. Teor. Fiz. \textbf{95}, 178 (1989),
\newblock [Sov. Phys. JETP \textbf{68}, 101 (1989)].

\bibitem{RosslerUK2006}
U.~K. R{\"o}{\ss}ler, A.~N. Bogdanov and C.~Pfleiderer,
\newblock \emph{Spontaneous skyrmion ground states in magnetic metals},
\newblock Nature \textbf{442}, 797 (2006),
\newblock \doi{10.1038/nature05056}.

\bibitem{MuhlbauerS2009}
S.~M{\"u}hlbauer, B.~Binz, F.~Jonietz, C.~Pfleiderer, A.~Rosch, A.~Neubauer,
  R.~Georgii and P.~B{\"o}ni,
\newblock \emph{Skyrmion {{Lattice}} in a {{Chiral Magnet}}},
\newblock Science \textbf{323}, 915 (2009),
\newblock \doi{10.1126/science.1166767}.

\bibitem{YuXZ2010}
X.~Z. Yu, Y.~Onose, N.~Kanazawa, J.~H. Park, J.~H. Han, Y.~Matsui, N.~Nagaosa
  and Y.~Tokura,
\newblock \emph{Real-space observation of a two-dimensional skyrmion crystal},
\newblock Nature \textbf{465}, 901 (2010),
\newblock \doi{10.1038/nature09124}.

\bibitem{YuXZ2011}
X.~Z. Yu, N.~Kanazawa, Y.~Onose, K.~Kimoto, W.~Z. Zhang, S.~Ishiwata, Y.~Matsui
  and Y.~Tokura,
\newblock \emph{Near room-temperature formation of a skyrmion crystal in
  thin-films of the helimagnet {{FeGe}}},
\newblock Nat. Mater. \textbf{10}, 106 (2011),
\newblock \doi{10.1038/nmat2916}.

\bibitem{SekiS2012}
S.~Seki, X.~Z. Yu, S.~Ishiwata and Y.~Tokura,
\newblock \emph{Observation of {{Skyrmions}} in a {{Multiferroic Material}}},
\newblock Science \textbf{336}, 198 (2012),
\newblock \doi{10.1126/science.1214143}.

\bibitem{AdamsT2012}
T.~Adams, A.~Chacon, M.~Wagner, A.~Bauer, G.~Brandl, B.~Pedersen, H.~Berger,
  P.~Lemmens and C.~Pfleiderer,
\newblock \emph{Long-{{Wavelength Helimagnetic Order}} and {{Skyrmion Lattice
  Phase}} in {{Cu}}{\textsubscript{2}}{{OSeO}}{\textsubscript{3}}},
\newblock Phys. Rev. Lett. \textbf{108}, 237204 (2012),
\newblock \doi{10.1103/PhysRevLett.108.237204}.

\bibitem{Dzyaloshinsky1958}
I.~Dzyaloshinsky,
\newblock \emph{A thermodynamic theory of weak ferromagnetism of
  antiferromagnetics},
\newblock J. Phys. Chem. Solids \textbf{4}, 241 (1958),
\newblock \doi{10.1016/0022-3697(58)90076-3}.

\bibitem{MoriyaT1960}
T.~Moriya,
\newblock \emph{New {{Mechanism}} of {{Anisotropic Superexchange
  Interaction}}},
\newblock Phys. Rev. Lett. \textbf{4}, 228 (1960),
\newblock \doi{10.1103/PhysRevLett.4.228}.

\bibitem{MoriyaT1960a}
T.~Moriya,
\newblock \emph{Anisotropic {{Superexchange Interaction}} and {{Weak
  Ferromagnetism}}},
\newblock Phys. Rev. \textbf{120}, 91 (1960),
\newblock \doi{10.1103/PhysRev.120.91}.

\bibitem{YuX2012}
X.~Yu, M.~Mostovoy, Y.~Tokunaga, W.~Zhang, K.~Kimoto, Y.~Matsui, Y.~Kaneko,
  N.~Nagaosa and Y.~Tokura,
\newblock \emph{Magnetic stripes and skyrmions with helicity reversals},
\newblock Proc. Natl. Acad. Sci. U.S.A. \textbf{109}, 8856 (2012),
\newblock \doi{10.1073/pnas.1118496109}.

\bibitem{YuXZ2014}
X.~Z. Yu, Y.~Tokunaga, Y.~Kaneko, W.~Z. Zhang, K.~Kimoto, Y.~Matsui, Y.~Taguchi
  and Y.~Tokura,
\newblock \emph{Biskyrmion states and their current-driven motion in a layered
  manganite},
\newblock Nat. Commun. \textbf{5}, 3198 (2014),
\newblock \doi{10.1038/ncomms4198}.

\bibitem{MallikR1998a}
R.~Mallik, E.~V. Sampathkumaran, P.~L. Paulose, H.~Sugawara and H.~Sato,
\newblock \emph{Magnetic anomalies in {{Gd}}{$_{2}$}{{PdSi}}{$_{3}$}},
\newblock Pramana - J. Phys. \textbf{51}, 505 (1998),
\newblock \doi{10.1007/BF02828942}.

\bibitem{SahaSR1999}
S.~R. Saha, H.~Sugawara, T.~D. Matsuda, H.~Sato, R.~Mallik and E.~V.
  Sampathkumaran,
\newblock \emph{Magnetic anisotropy, first-order-like metamagnetic transitions,
  and large negative magnetoresistance in single-crystal
  {{Gd}}{$_{2}$}{{PdSi}}{$_{3}$}},
\newblock Phys. Rev. B \textbf{60}, 12162 (1999),
\newblock \doi{10.1103/PhysRevB.60.12162}.

\bibitem{KurumajiT2019}
T.~Kurumaji, T.~Nakajima, M.~Hirschberger, A.~Kikkawa, Y.~Yamasaki,
  H.~Sagayama, H.~Nakao, Y.~Taguchi, T.-h. Arima and Y.~Tokura,
\newblock \emph{Skyrmion lattice with a giant topological {{Hall}} effect in a
  frustrated triangular-lattice magnet},
\newblock Science \textbf{365}, 914 (2019),
\newblock \doi{10.1126/science.aau0968}.

\bibitem{ChandragiriV2016}
V.~Chandragiri, K.~K. Iyer and E.~V. Sampathkumaran,
\newblock \emph{Magnetic behavior of
  {{Gd}}{\textsubscript{3}}{{Ru}}{\textsubscript{4}}{{Al}}{\textsubscript{12}},
  a layered compound with distorted kagom\'e net},
\newblock J. Phys.: Condens. Matter \textbf{28}, 286002 (2016),
\newblock \doi{10.1088/0953-8984/28/28/286002}.

\bibitem{HirschbergerM2019}
M.~Hirschberger, T.~Nakajima, S.~Gao, L.~Peng, A.~Kikkawa, T.~Kurumaji,
  M.~Kriener, Y.~Yamasaki, H.~Sagayama, H.~Nakao, K.~Ohishi, K.~Kakurai
  \emph{et~al.},
\newblock \emph{Skyrmion phase and competing magnetic orders on a breathing
  kagom\'e lattice},
\newblock Nat. Commun. \textbf{10}, 5831 (2019),
\newblock \doi{10.1038/s41467-019-13675-4}.

\bibitem{KhanhND2020}
N.~D. Khanh, T.~Nakajima, X.~Yu, S.~Gao, K.~Shibata, M.~Hirschberger,
  Y.~Yamasaki, H.~Sagayama, H.~Nakao, L.~Peng, K.~Nakajima, R.~Takagi
  \emph{et~al.},
\newblock \emph{Nanometric square skyrmion lattice in a centrosymmetric
  tetragonal magnet},
\newblock Nat. Nanotechnol. \textbf{15}, 444 (2020),
\newblock \doi{10.1038/s41565-020-0684-7}.

\bibitem{AmorosoD2020}
D.~Amoroso, P.~Barone and S.~Picozzi,
\newblock \emph{Spontaneous skyrmionic lattice from anisotropic symmetric
  exchange in a {{Ni-halide}} monolayer},
\newblock Nat. Commun. \textbf{11}, 5784 (2020),
\newblock \doi{10.1038/s41467-020-19535-w}.

\bibitem{ChakrabarttyD2022}
D.~Chakrabartty, S.~Jamaluddin, S.~K. Manna and A.~K. Nayak,
\newblock \emph{Tunable room temperature magnetic skyrmions in centrosymmetric
  kagome magnet {{Mn}}{$_{4}$}{{Ga}}{$_{2}$}{{Sn}}},
\newblock Commun. Phys. \textbf{5}, 1 (2022),
\newblock \doi{10.1038/s42005-022-00971-7}.

\bibitem{ShangT2021}
T.~Shang, Y.~Xu, D.~J. Gawryluk, J.~Z. Ma, T.~Shiroka, M.~Shi and
  E.~Pomjakushina,
\newblock \emph{Anomalous {{Hall}} resistivity and possible topological
  {{Hall}} effect in the {{EuAl}}{\textsubscript{4}} antiferromagnet},
\newblock Phys. Rev. B \textbf{103}, L020405 (2021),
\newblock \doi{10.1103/PhysRevB.103.L020405}.

\bibitem{KanekoK2021}
K.~Kaneko, T.~Kawasaki, A.~Nakamura, K.~Munakata, A.~Nakao, T.~Hanashima,
  R.~Kiyanagi, T.~Ohhara, M.~Hedo, T.~Nakama and Y.~{\=O}nuki,
\newblock \emph{Charge-{{Density-Wave Order}} and {{Multiple Magnetic
  Transitions}} in {{Divalent Europium Compound EuAl}}{\textsubscript{4}}},
\newblock J. Phys. Soc. Jpn. \textbf{90}, 064704 (2021),
\newblock \doi{10.7566/JPSJ.90.064704}.

\bibitem{TakagiR2022}
R.~Takagi, N.~Matsuyama, V.~Ukleev, L.~Yu, J.~S. White, S.~Francoual, J.~R.~L.
  Mardegan, S.~Hayami, H.~Saito, K.~Kaneko, K.~Ohishi, Y.~{\=O}nuki
  \emph{et~al.},
\newblock \emph{Square and rhombic lattices of magnetic skyrmions in a
  centrosymmetric binary compound},
\newblock Nat. Commun. \textbf{13}, 1472 (2022),
\newblock \doi{10.1038/s41467-022-29131-9}.

\bibitem{OkuboT2012}
T.~Okubo, S.~Chung and H.~Kawamura,
\newblock \emph{Multiple-q {{States}} and the {{Skyrmion Lattice}} of the
  {{Triangular-Lattice Heisenberg Antiferromagnet}} under {{Magnetic Fields}}},
\newblock Phys. Rev. Lett. \textbf{108}, 017206 (2012),
\newblock \doi{10.1103/PhysRevLett.108.017206}.

\bibitem{LeonovAO2015}
A.~O. Leonov and M.~Mostovoy,
\newblock \emph{Multiply periodic states and isolated skyrmions in an
  anisotropic frustrated magnet},
\newblock Nat. Commun. \textbf{6}, 8275 (2015),
\newblock \doi{10.1038/ncomms9275}.

\bibitem{LinSZ2016}
S.-Z. Lin and S.~Hayami,
\newblock \emph{Ginzburg-{{Landau}} theory for skyrmions in inversion-symmetric
  magnets with competing interactions},
\newblock Phys. Rev. B \textbf{93}, 064430 (2016),
\newblock \doi{10.1103/PhysRevB.93.064430}.

\bibitem{Hayami16}
S.~Hayami, S.-Z. Lin and C.~D. Batista,
\newblock \emph{Bubble and skyrmion crystals in frustrated magnets with
  easy-axis anisotropy},
\newblock Phys. Rev. B \textbf{93}, 184413 (2016),
\newblock \doi{10.1103/PhysRevB.93.184413}.

\bibitem{BatistaCD2016_review}
C.~D. Batista, S.-Z. Lin, S.~Hayami and Y.~Kamiya,
\newblock \emph{Frustration and chiral orderings in correlated electron
  systems},
\newblock Rep. Prog. Phys. \textbf{79}, 084504 (2016),
\newblock \doi{10.1088/0034-4885/79/8/084504}.

\bibitem{Zhang20}
S.-S. Zhang, H.~Ishizuka, H.~Zhang, G.~B. Hal\'asz and C.~D. Batista,
\newblock \emph{Real-space berry curvature of itinerant electron systems with
  spin-orbit interaction},
\newblock Phys. Rev. B \textbf{101}, 024420 (2020),
\newblock \doi{10.1103/PhysRevB.101.024420}.

\bibitem{OnodaM2004}
M.~Onoda, G.~Tatara and N.~Nagaosa,
\newblock \emph{Anomalous {{Hall Effect}} and {{Skyrmion Number}} in {{Real}}
  and {{Momentum Spaces}}},
\newblock J. Phys. Soc. Jpn. \textbf{73}, 2624 (2004),
\newblock \doi{10.1143/JPSJ.73.2624}.

\bibitem{YiSD2009}
S.~D. Yi, S.~Onoda, N.~Nagaosa and J.~H. Han,
\newblock \emph{Skyrmions and anomalous {{Hall}} effect in a
  {{Dzyaloshinskii-Moriya}} spiral magnet},
\newblock Phys. Rev. B \textbf{80}, 054416 (2009),
\newblock \doi{10.1103/PhysRevB.80.054416}.

\bibitem{HamamotoK2015}
K.~Hamamoto, M.~Ezawa and N.~Nagaosa,
\newblock \emph{Quantized topological {{Hall}} effect in skyrmion crystal},
\newblock Phys. Rev. B \textbf{92}, 115417 (2015),
\newblock \doi{10.1103/PhysRevB.92.115417}.

\bibitem{GobelB2017}
B.~G{\"o}bel, A.~Mook, J.~Henk and I.~Mertig,
\newblock \emph{Unconventional topological {{Hall}} effect in skyrmion crystals
  caused by the topology of the lattice},
\newblock Phys. Rev. B \textbf{95}, 094413 (2017),
\newblock \doi{10.1103/PhysRevB.95.094413}.

\bibitem{JonietzF2010}
F.~Jonietz, S.~M{\"u}hlbauer, C.~Pfleiderer, A.~Neubauer, W.~M{\"u}nzer,
  A.~Bauer, T.~Adams, R.~Georgii, P.~B{\"o}ni, R.~A. Duine, K.~Everschor,
  M.~Garst \emph{et~al.},
\newblock \emph{Spin {{Transfer Torques}} in {{MnSi}} at {{Ultralow Current
  Densities}}},
\newblock Science \textbf{330}, 1648 (2010),
\newblock \doi{10.1126/science.1195709}.

\bibitem{YuXZ2012}
X.~Z. Yu, N.~Kanazawa, W.~Z. Zhang, T.~Nagai, T.~Hara, K.~Kimoto, Y.~Matsui,
  Y.~Onose and Y.~Tokura,
\newblock \emph{Skyrmion flow near room temperature in an ultralow current
  density},
\newblock Nat. Commun. \textbf{3}, 988 (2012),
\newblock \doi{10.1038/ncomms1990}.

\bibitem{SchulzT2012}
T.~Schulz, R.~Ritz, A.~Bauer, M.~Halder, M.~Wagner, C.~Franz, C.~Pfleiderer,
  K.~Everschor, M.~Garst and A.~Rosch,
\newblock \emph{Emergent electrodynamics of skyrmions in a chiral magnet},
\newblock Nat. Phys. \textbf{8}, 301 (2012),
\newblock \doi{10.1038/nphys2231}.

\bibitem{NagaosaN2013}
N.~Nagaosa and Y.~Tokura,
\newblock \emph{Topological properties and dynamics of magnetic skyrmions},
\newblock Nat. Nanotechnol. \textbf{8}, 899 (2013),
\newblock \doi{10.1038/nnano.2013.243}.

\bibitem{WangZ2020}
Z.~Wang, Y.~Su, S.-Z. Lin and C.~D. Batista,
\newblock \emph{Skyrmion {{Crystal}} from {{RKKY Interaction Mediated}} by {{2D
  Electron Gas}}},
\newblock Phys. Rev. Lett. \textbf{124}, 207201 (2020),
\newblock \doi{10.1103/PhysRevLett.124.207201}.

\bibitem{RudermanMA1954}
M.~A. Ruderman and C.~Kittel,
\newblock \emph{Indirect {{Exchange Coupling}} of {{Nuclear Magnetic Moments}}
  by {{Conduction Electrons}}},
\newblock Phys. Rev. \textbf{96}, 99 (1954),
\newblock \doi{10.1103/PhysRev.96.99}.

\bibitem{KasuyaT1956}
T.~Kasuya,
\newblock \emph{A {{Theory}} of {{Metallic Ferro-}} and {{Antiferromagnetism}}
  on {{Zener}}'s {{Model}}},
\newblock Prog. Theor. Phys. \textbf{16}, 45 (1956),
\newblock \doi{10.1143/PTP.16.45}.

\bibitem{YosidaK1957}
K.~Yosida,
\newblock \emph{Magnetic {{Properties}} of {{Cu-Mn Alloys}}},
\newblock Phys. Rev. \textbf{106}, 893 (1957),
\newblock \doi{10.1103/PhysRev.106.893}.

\bibitem{MartinI2008}
I.~Martin and C.~D. Batista,
\newblock \emph{Itinerant {{Electron-Driven Chiral Magnetic Ordering}} and
  {{Spontaneous Quantum Hall Effect}} in {{Triangular Lattice Models}}},
\newblock Phys. Rev. Lett. \textbf{101}, 156402 (2008),
\newblock \doi{10.1103/PhysRevLett.101.156402}.

\bibitem{park2023}
P.~Park, W.~Cho, C.~Kim, Y.~An, Y.-G. Kang, M.~Avdeev, R.~Sibille, K.~Iida,
  R.~Kajimoto, K.~H. Lee, W.~Ju, E.-J. Cho \emph{et~al.},
\newblock \emph{Tetrahedral triple-q ordering in the metallic triangular
  lattice antiferromagnet co1/3tas2} (2023), \eprint{2303.03760}.

\bibitem{AkagiY2012}
Y.~Akagi, M.~Udagawa and Y.~Motome,
\newblock \emph{Hidden {{Multiple-Spin Interactions}} as an {{Origin}} of
  {{Spin Scalar Chiral Order}} in {{Frustrated Kondo Lattice Models}}},
\newblock Phys. Rev. Lett. \textbf{108}, 096401 (2012),
\newblock \doi{10.1103/PhysRevLett.108.096401}.

\bibitem{OzawaR2017}
R.~Ozawa, S.~Hayami and Y.~Motome,
\newblock \emph{Zero-{{Field Skyrmions}} with a {{High Topological Number}} in
  {{Itinerant Magnets}}},
\newblock Phys. Rev. Lett. \textbf{118}, 147205 (2017),
\newblock \doi{10.1103/PhysRevLett.118.147205}.

\bibitem{HayamiS2021_review}
S.~Hayami and Y.~Motome,
\newblock \emph{Topological spin crystals by itinerant frustration},
\newblock J. Phys.: Condens. Matter \textbf{33}, 443001 (2021),
\newblock \doi{10.1088/1361-648X/ac1a30}.

\bibitem{HayamiS2017}
S.~Hayami, R.~Ozawa and Y.~Motome,
\newblock \emph{Effective bilinear-biquadratic model for noncoplanar ordering
  in itinerant magnets},
\newblock Phys. Rev. B \textbf{95}, 224424 (2017),
\newblock \doi{10.1103/PhysRevB.95.224424}.

\bibitem{Kurz01}
P.~Kurz, G.~Bihlmayer, K.~Hirai and S.~Bl\"ugel,
\newblock \emph{Three-{{Dimensional Spin Structure}} on a {{Two-Dimensional
  Lattice}}: {{Mn}}/{{Cu}}(111)},
\newblock Phys. Rev. Lett. \textbf{86}, 1106 (2001),
\newblock \doi{10.1103/PhysRevLett.86.1106}.

\bibitem{HeinzeS2011}
S.~Heinze, K.~{von Bergmann}, M.~Menzel, J.~Brede, A.~Kubetzka,
  R.~Wiesendanger, G.~Bihlmayer and S.~Bl{\"u}gel,
\newblock \emph{Spontaneous atomic-scale magnetic skyrmion lattice in two
  dimensions},
\newblock Nat. Phys. \textbf{7}, 713 (2011),
\newblock \doi{10.1038/nphys2045}.

\bibitem{KumarS2010}
S.~Kumar and J.~{van den Brink},
\newblock \emph{Frustration-{{Induced Insulating Chiral Spin State}} in
  {{Itinerant Triangular-Lattice Magnets}}},
\newblock Phys. Rev. Lett. \textbf{105}, 216405 (2010),
\newblock \doi{10.1103/PhysRevLett.105.216405}.

\bibitem{RejaS2015}
S.~Reja, R.~Ray, J.~{van den Brink} and S.~Kumar,
\newblock \emph{Coupled spin-charge order in frustrated itinerant triangular
  magnets},
\newblock Phys. Rev. B \textbf{91}, 140403 (2015),
\newblock \doi{10.1103/PhysRevB.91.140403}.

\bibitem{KathyatDS2020}
D.~S. Kathyat, A.~Mukherjee and S.~Kumar,
\newblock \emph{Microscopic magnetic {{Hamiltonian}} for exotic spin textures
  in metals},
\newblock Phys. Rev. B \textbf{102}, 075106 (2020),
\newblock \doi{10.1103/PhysRevB.102.075106}.

\bibitem{KathyatDS2021}
D.~S. Kathyat, A.~Mukherjee and S.~Kumar,
\newblock \emph{Electronic mechanism for nanoscale skyrmions and topological
  metals},
\newblock Phys. Rev. B \textbf{103}, 035111 (2021),
\newblock \doi{10.1103/PhysRevB.103.035111}.

\bibitem{MitsumotoK2022}
K.~Mitsumoto and H.~Kawamura,
\newblock \emph{Skyrmion crystal in the {{RKKY}} system on the two-dimensional
  triangular lattice},
\newblock Phys. Rev. B \textbf{105}, 094427 (2022),
\newblock \doi{10.1103/PhysRevB.105.094427}.

\bibitem{WangZ2016a}
Z.~Wang, K.~Barros, G.-W. Chern, D.~L. Maslov and C.~D. Batista,
\newblock \emph{Resistivity {{Minimum}} in {{Highly Frustrated Itinerant
  Magnets}}},
\newblock Phys. Rev. Lett. \textbf{117}, 206601 (2016),
\newblock \doi{10.1103/PhysRevLett.117.206601}.

\bibitem{YunokiS1998}
S.~Yunoki, J.~Hu, A.~L. Malvezzi, A.~Moreo, N.~Furukawa and E.~Dagotto,
\newblock \emph{Phase {{Separation}} in {{Electronic Models}} for
  {{Manganites}}},
\newblock Phys. Rev. Lett. \textbf{80}, 845 (1998),
\newblock \doi{10.1103/PhysRevLett.80.845}.

\bibitem{WeisseA2006_RMP}
A.~Wei{\ss}e, G.~Wellein, A.~Alvermann and H.~Fehske,
\newblock \emph{The kernel polynomial method},
\newblock Rev. Mod. Phys. \textbf{78}, 275 (2006),
\newblock \doi{10.1103/RevModPhys.78.275}.

\bibitem{MotomeY1999}
Y.~Motome and N.~Furukawa,
\newblock \emph{A {{Monte Carlo Method}} for {{Fermion Systems}} {{Coupled}}
  with {{Classical Degrees}} of {{Freedom}}},
\newblock J. Phys. Soc. Jpn. \textbf{68}, 3853 (1999),
\newblock \doi{10.1143/JPSJ.68.3853}.

\bibitem{MotomeY2000}
Y.~Motome and N.~Furukawa,
\newblock \emph{Critical {{Temperature}} of {{Ferromagnetic Transition}} in
  {{Three-Dimensional Double-Exchange Models}}},
\newblock J. Phys. Soc. Jpn. \textbf{69}, 3785 (2000),
\newblock \doi{10.1143/JPSJ.69.3785}.

\bibitem{BarrosK2013}
K.~Barros and Y.~Kato,
\newblock \emph{Efficient {{Langevin}} simulation of coupled classical fields
  and fermions},
\newblock Phys. Rev. B \textbf{88}, 235101 (2013),
\newblock \doi{10.1103/PhysRevB.88.235101}.

\bibitem{TangJM2012}
J.~M. Tang and Y.~Saad,
\newblock \emph{A probing method for computing the diagonal of a matrix
  inverse},
\newblock Numer. Linear Algebra Appl. \textbf{19}, 485 (2012),
\newblock \doi{10.1002/nla.779}.

\bibitem{WangZ2018}
Z.~Wang, G.-W. Chern, C.~D. Batista and K.~Barros,
\newblock \emph{Gradient-based stochastic estimation of the density matrix},
\newblock J. Chem. Phys. \textbf{148}, 094107 (2018),
\newblock \doi{10.1063/1.5017741}.

\bibitem{SolenovD2012}
D.~Solenov, D.~Mozyrsky and I.~Martin,
\newblock \emph{Chirality {{Waves}} in {{Two-Dimensional Magnets}}},
\newblock Phys. Rev. Lett. \textbf{108}, 096403 (2012),
\newblock \doi{10.1103/PhysRevLett.108.096403}.

\bibitem{OzawaR2016}
R.~Ozawa, S.~Hayami, K.~Barros, G.-W. Chern, Y.~Motome and C.~D. Batista,
\newblock \emph{Vortex {{Crystals}} with {{Chiral Stripes}} in {{Itinerant
  Magnets}}},
\newblock J. Phys. Soc. Jpn. \textbf{85}, 103703 (2016),
\newblock \doi{10.7566/JPSJ.85.103703}.

\bibitem{KamiyaY2014}
Y.~Kamiya and C.~D. Batista,
\newblock \emph{Magnetic {{Vortex Crystals}} in {{Frustrated Mott Insulator}}},
\newblock Phys. Rev. X \textbf{4}, 011023 (2014),
\newblock \doi{10.1103/PhysRevX.4.011023}.

\bibitem{WangZ2015}
Z.~Wang, Y.~Kamiya, A.~H. Nevidomskyy and C.~D. Batista,
\newblock \emph{Three-{{Dimensional Crystallization}} of {{Vortex Strings}} in
  {{Frustrated Quantum Magnets}}},
\newblock Phys. Rev. Lett. \textbf{115}, 107201 (2015),
\newblock \doi{10.1103/PhysRevLett.115.107201}.

\bibitem{WangZ2021}
Z.~Wang, Y.~Su, S.-Z. Lin and C.~D. Batista,
\newblock \emph{Meron, skyrmion, and vortex crystals in centrosymmetric
  tetragonal magnets},
\newblock Phys. Rev. B \textbf{103}, 104408 (2021),
\newblock \doi{10.1103/PhysRevB.103.104408}.

\bibitem{AkagiY2010}
Y.~Akagi and Y.~Motome,
\newblock \emph{Spin {{Chirality Ordering}} and {{Anomalous Hall Effect}} in
  the {{Ferromagnetic Kondo Lattice Model}} on a {{Triangular Lattice}}},
\newblock J. Phys. Soc. Jpn. \textbf{79}, 083711 (2010),
\newblock \doi{10.1143/JPSJ.79.083711}.

\bibitem{KatoY2010}
Y.~Kato, I.~Martin and C.~D. Batista,
\newblock \emph{Stability of the {{Spontaneous Quantum Hall State}} in the
  {{Triangular Kondo-Lattice Model}}},
\newblock Phys. Rev. Lett. \textbf{105}, 266405 (2010),
\newblock \doi{10.1103/PhysRevLett.105.266405}.

\bibitem{BarrosK2014}
K.~Barros, J.~W.~F. Venderbos, G.-W. Chern and C.~D. Batista,
\newblock \emph{Exotic magnetic orderings in the kagome {{Kondo-lattice}}
  model},
\newblock Phys. Rev. B \textbf{90}, 245119 (2014),
\newblock \doi{10.1103/PhysRevB.90.245119}.

\bibitem{XuY2021}
Y.~Xu, L.~Das, J.~Z. Ma, C.~J. Yi, S.~M. Nie, Y.~G. Shi, A.~Tiwari, S.~S.
  Tsirkin, T.~Neupert, M.~Medarde, M.~Shi, J.~Chang \emph{et~al.},
\newblock \emph{Unconventional {{Transverse Transport}} above and below the
  {{Magnetic Transition Temperature}} in {{Weyl Semimetal
  EuCd}}{\textsubscript{2}}{{As}}{\textsubscript{2}}},
\newblock Phys. Rev. Lett. \textbf{126}, 076602 (2021),
\newblock \doi{10.1103/PhysRevLett.126.076602}.

\bibitem{SeoS2021}
S.~Seo, S.~Hayami, Y.~Su, S.~M. Thomas, F.~Ronning, E.~D. Bauer, J.~D.
  Thompson, S.-Z. Lin and P.~F.~S. Rosa,
\newblock \emph{Spin-texture-driven electrical transport in multi-{{Q}}
  antiferromagnets},
\newblock Commun. Phys. \textbf{4}, 1 (2021),
\newblock \doi{10.1038/s42005-021-00558-8}.

\bibitem{ZhangH2021a}
H.~Zhang, X.~Y. Zhu, Y.~Xu, D.~J. Gawryluk, W.~Xie, S.~L. Ju, M.~Shi,
  T.~Shiroka, Q.~F. Zhan, E.~Pomjakushina and T.~Shang,
\newblock \emph{Giant magnetoresistance and topological {{Hall}} effect in the
  {{EuGa}}{\textsubscript{4}} antiferromagnet},
\newblock J. Phys.: Condens. Matter \textbf{34}, 034005 (2021),
\newblock \doi{10.1088/1361-648X/ac3102}.

\bibitem{MoyaJM2022}
J.~M. Moya, S.~Lei, E.~M. Clements, C.~S. Kengle, S.~Sun, K.~Allen, Q.~Li,
  Y.~Y. Peng, A.~A. Husain, M.~Mitrano, M.~J. Krogstad, R.~Osborn
  \emph{et~al.},
\newblock \emph{Incommensurate magnetic orders and topological {{Hall}} effect
  in the square-net centrosymmetric
  {{EuGa}}{\textsubscript{2}}{{Al}}{\textsubscript{2}} system},
\newblock Phys. Rev. Materials \textbf{6}, 074201 (2022),
\newblock \doi{10.1103/PhysRevMaterials.6.074201}.

\bibitem{WangW2016}
W.~Wang, Y.~Zhang, G.~Xu, L.~Peng, B.~Ding, Y.~Wang, Z.~Hou, X.~Zhang, X.~Li,
  E.~Liu, S.~Wang, J.~Cai \emph{et~al.},
\newblock \emph{A {{Centrosymmetric Hexagonal Magnet}} with {{Superstable
  Biskyrmion Magnetic Nanodomains}} in a {{Wide Temperature Range}} of
  100\textendash 340 {{K}}},
\newblock Advanced Materials \textbf{28}, 6887 (2016),
\newblock \doi{10.1002/adma.201600889}.

\bibitem{LiH2019}
H.~Li, B.~Ding, J.~Chen, Z.~Li, Z.~Hou, E.~Liu, H.~Zhang, X.~Xi, G.~Wu and
  W.~Wang,
\newblock \emph{Large topological {{Hall}} effect in a geometrically frustrated
  kagome magnet {{Fe}}{\textsubscript{3}}{{Sn}}{\textsubscript{2}}},
\newblock Appl. Phys. Lett. \textbf{114}, 192408 (2019),
\newblock \doi{10.1063/1.5088173}.

\bibitem{WangS2020}
S.~Wang, Q.~Zeng, D.~Liu, H.~Zhang, L.~Ma, G.~Xu, Y.~Liang, Z.~Zhang, H.~Wu,
  R.~Che, X.~Han and Q.~Huang,
\newblock \emph{Giant {{Topological Hall Effect}} and {{Superstable Spontaneous
  Skyrmions}} below 330 {{K}} in a {{Centrosymmetric Complex Noncollinear
  Ferromagnet NdMn}}{\textsubscript{2}}{{Ge}}{\textsubscript{2}}},
\newblock ACS Appl. Mater. Interfaces \textbf{12}, 24125 (2020),
\newblock \doi{10.1021/acsami.0c04632}.

\bibitem{ZhengX2021}
X.~Zheng, X.~Zhao, J.~Qi, X.~Luo, S.~Ma, C.~Chen, H.~Zeng, G.~Yu, N.~Fang,
  S.~U. Rehman, W.~Ren, B.~Li \emph{et~al.},
\newblock \emph{Giant topological {{Hall}} effect around room temperature in
  noncollinear ferromagnet {{NdMn}}{\textsubscript{2}}{{Ge}}{\textsubscript{2}}
  single crystal},
\newblock Appl. Phys. Lett. \textbf{118}, 072402 (2021),
\newblock \doi{10.1063/5.0033379}.

\bibitem{DallyRL2021}
R.~L. Dally, J.~W. Lynn, N.~J. Ghimire, D.~Michel, P.~Siegfried and I.~I.
  Mazin,
\newblock \emph{Chiral properties of the zero-field spiral state and
  field-induced magnetic phases of the itinerant kagome metal
  {{YMn}}{\textsubscript{6}}{{Sn}}{\textsubscript{6}}},
\newblock Phys. Rev. B \textbf{103}, 094413 (2021),
\newblock \doi{10.1103/PhysRevB.103.094413}.

\bibitem{NabiMRU2021}
M.~R.~U. Nabi, A.~Wegner, F.~Wang, Y.~Zhu, Y.~Guan, A.~Fereidouni, K.~Pandey,
  R.~Basnet, G.~Acharya, H.~O.~H. Churchill, Z.~Mao and J.~Hu,
\newblock \emph{Giant topological {{Hall}} effect in centrosymmetric tetragonal
  {{Mn}}{\textsubscript{2-x}}{{Zn}}{\textsubscript{x}}{{Sb}}},
\newblock Phys. Rev. B \textbf{104}, 174419 (2021),
\newblock \doi{10.1103/PhysRevB.104.174419}.

\bibitem{SenechalD2008}
D.~S{\'e}n{\'e}chal,
\newblock \emph{An introduction to quantum cluster methods}
  \eprint{http://arxiv.org/abs/0806.2690}.

\bibitem{nlopt}
S.~G. Johnson,
\newblock \emph{The {{NLopt}} nonlinear-optimization package}
  \eprint{http://github.com/stevengj/nlopt}.

\bibitem{NocedalJ1980}
J.~Nocedal,
\newblock \emph{Updating {{Quasi-Newton Matrices}} with {{Limited Storage}}},
\newblock Math. Comp. \textbf{35}, 773 (1980),
\newblock \doi{10.1090/S0025-5718-1980-0572855-7}.

\bibitem{LiuDC1989}
D.~C. Liu and J.~Nocedal,
\newblock \emph{On the limited memory {{BFGS}} method for large scale
  optimization},
\newblock Mathematical Programming \textbf{45}, 503 (1989),
\newblock \doi{10.1007/BF01589116}.

\end{thebibliography}

\nolinenumbers

\end{document}